# Value, Variety and Viability:
# New Business Models for Co-Creation in Outcome-based Contracts

*Irene Ng[1], University of Warwick, United Kingdom*

*Gerard Briscoe[2], University of Cambridge, United Kingdom*

**ABSTRACT**

*We propose that designing a manufacturer's equipment-based service value proposition in outcome-based contracts is the design of a new business model capable of managing threats to the firm's viability that can arise from the contextual variety of use that customers may subject the firm's value propositions. Furthermore, manufacturers need to understand these emerging business models as the capability of managing both asset and service provision to achieve use outcomes with customers, including emotional outcomes such as customer experience. Service-Dominant Logic proposes that all "goods are a distribution mechanism for service provision", upon which we propose a value-centric approach to understanding the interactions between the asset and service provision, and suggest a viable systems approach towards reorganising the firm to achieve such a business model. Three case studies of B2B equipment-based service systems were analysed to understand customers' co-creation activities in achieving outcomes, in which we found that the co-creation of complex multi-dimensional value could be delivered through the different value propositions of the firm catering to different aspects (dimensions) of the value to be co-created. The study provides a way for managers to understand the effectiveness (rather than efficiency) of firms in adopting emerging business models that design for value co-creation in what are ultimately complex socio-technical systems.*

---

[1] *Irene CL Ng is Professor of Marketing and Service Systems with the Service Systems Group at Warwick Manufacturing Group in the University of Warwick. E-mail: irene.ng@warwick.ac.uk*

[2] *Gerard Briscoe is a Research Associate at the Systems Research Group, Computing Laboratory, University of Cambridge. E-mail: gerard.briscoe@cl.cam.ac.uk*

# INTRODUCTION

While manufacturing in the past century has been essential to wealth creation, developed economies are gradually becoming service-oriented (Ramirez, 1999). Research recommends that manufacturers should diversify into providing services to remain viable, aiming to facilitate equipment use for customer outcomes rather than just transferring the ownership of equipment ( Neely, 2008; Baines et al, 2007). This means that the value proposition of the manufacturer changes from *exchange value* obtained from equipment provision, to *value-in-use*, obtained from the outcomes of equipment *use.* Outcome-based contracts such as Rolls-Royce's "Power-by-the-hour®", exemplifies such a change in value proposition, as the firm is paid not according to its service activities such as material and repairs, but based on the outcome of such activities in continual use situations i.e. the number of hours of engine in the air. This change in business model requires firm-customer relationships to be embedded in the processes and interactions of collaborative value-creating activities, ie value co-creation. Therefore, cooperation between the firm and its customer is a partnership that requires a "mutual and synergistic pooling of resources and capabilities and a substantial degree of co-mingling between partners in terms of people, systems, skills etc. in order to attain their objectives" (Madhok & Tallman, 1998).

Given the challenge of having to design a manufacturer's value propositions for more effective collaboration with their customers, we suggest that this can be best understood through the conceptualisation of service proposed by the Service-Dominant (S-D) Logic (Vargo & Lusch, 2004, 2008), where assets (goods) are seen to be indirect service provision. Through a S-D Logic approach, we propose three key issues for the understanding of outcome-based contracts as a new business model.

First, manufacturers must understand the interactions between asset and human activities provision when combined as value propositions, and what is the intended **value** to be co-created for customer outcomes.

Second, a comprehension of value-in-use also requires the understanding of contexts in which value creation occurs. The greater the **variety** of contexts, the greater could be the challenge in design, due to the increased complexity that can arise from supporting the system under contextual variety. This becomes most acute for outcome-based contracts, since continual use of equipment sits within the customer's space and requires the customer's resources to achieve use for their own goal, increasing the variety.

Finally, since contextual variety of use will impact upon the firm's value propositions, achieving outcomes of use as part of contract performance can become increasingly complex, even threatening the firm's future profitability and continued **viability**. Therefore, firms need to re-organise themselves to maintain viability, and manage the complexity that can emerge from such service systems. We propose a viable systems approach, which provides a model of organisation for the firm to maintain viability. We consider firms transitioning from being a manufacturer to a system of achieving value-in-use in co-creation with their customer under this approach, and analyse three longitudinal case studies of manufacturers moving to outcome-based service provision over a three-year period.

We found the nature of value to be co-created to be beyond the functional and to include the emotional, i.e. the customer experience. Second, the degree of contextual variety threatens the stability of the system and finally, the firm's 'legacy' viability is seen as a challenge in achieving co-creation. To counter the viability threat, the firm (a) uses Asset Provision for Scalability and Replicability of the value proposition and (b) Human Activities Provision for variety absorption and co-creating emotional value (customer experience), and (c) manages the resources of the customer in achieving outcomes with the firm to improve the scalability and stability of the firm's provision. Overall, the firms came to realise that an asset was not exogenous to the service system and that it could be redesigned to absorb contextual variety of use, which would then impact on the effectiveness of human activities for service provisioning, enabling the firm to scale and replicate the provisioning across contracts. Furthermore, our study suggests that organisations structured around manufacturing require a re-evaluation of their operational elements and viability when they transform into hybrid manufacturing-service organisations. We argue for a transformation in the customer relationship to help realise the value proposition that firms offer. Specifically, we propose a viable systems approach for the inclusion of customer activities within the firm's boundaries of management and operation for value co-creation, and our paper argues how this could be achieved while maintaining viability.

The remainder of the paper is organised as follows. A literature review considering the theoretical links between value, variety and viability in designing for value co-creation in complex service systems is presented. This is followed by the methodology for the longitudinal case studies of manufacturers contracting based on outcomes of equipment, compelling a value co-creation approach. The findings from these case studies are then used to address the research question of threats to viability from value co-creation under contextual variety. We then discuss an extension of the S-D Logic approach for organising the firm through viable systems. We conclude with the managerial implications on this new way of configuring the organisation for effectiveness, designing for value co-creation in what are ultimately complex socio-technical systems.

## LITERATURE REVIEW

### Business Models

Since the rise of the Internet and proliferation of e-business in the 1990s, the business model concept has been increasingly discussed in academic literature. However, over the last two decades it has become clear that research into business models includes very different perspectives. In management studies, it has grown independently within the different management disciplines, with little cross-disciplinary understanding (Zott, Amit & Massa, 2011). Still, most of these literature agree that business models comprise key aspects of different elements, with the most frequent mentions being "economic model", "target markets", "firms value offering", "partner network and roles", "customer interface/relationship" and "internal infrastructure/connected activities" (Morris, Schindehutte & Allen, 2005).

Correspondingly, we find numerous definitions for business models. Zott and Amit (2007) consider it as "the structure, content, and governance of transactions between the focal firm and its exchange partners, and represents a conceptualisation of the pattern of transactional links between the firm and its exchange partners". Shafer, Smith, and Linder (2005) define it as "a representation of a firm's underlying core logic and strategic choices for creating and capturing value within a value network". Others include a "system manifested in the components and related material and cognitive aspects comprising key components including the company's network of relationships, operations and resource base" (Tikkanen, Lamberg, Parvinen & Kallunki, 2005), a "construct that mediates the value creation process" (Chesbrough & Rosenbloom, 2002), and "configurations of interrelated capabilities, governing the content, process and management of the interaction and exchange in dyadic value co-creation" (Storbacka & Nenonen, 2009).

Studies into business models have endeavoured to define common themes across these different meanings. Shafer et al (2005) suggest classification into four primary components; (1) strategic choices, (2) the value network, (3) creating value, and (4) capturing value. Baden-Fuller and Morgan (2010) suggest three approaches to studying business models. First, scale models (taxonomy) and role models (typology), where successful firms shape their industries, inspiring imitation and therefore encouraging their own further innovation. Second, study as an organism model in biology, including systems thinking for understanding how knowledge is created (Creager, Lunbeck & Norton, 2007). Finally, as a "portfolio" of elements to create a successful business, Zott et al (2011) highlight four emerging common themes: (1) the business model should be the unit of analysis instead of its component parts, (2) the need for system-level thinking because dynamic activities are performed by the firm and by third parties, (3) organisational activities play a critical role and (4) business models explain how value is captured and created at different levels of the organisation (as well as the different stakeholders).

The varied definitions and studies considered strongly suggest that new business models occur primarily from innovation or new technology. So, firms needed to alter their strategies to meet new challenges. Indeed, we consider a change in business model as the ability to identify different value drivers of the business, and changing where necessary to build and maintain sustainable performance over time. Furthermore, we propose four common themes to business model studies. First, new business models often result from changes in value drivers. Second, firms can improve competitive advantage and performance through changes in such value drivers. Third, network or partnership studies features prominently in business model literature (Zott & Amit, 2009; Johnson, Christensen & Kagermann, 2008; Magretta, 2002; Demil & Lecocq, 2010). Fourth, focus on new business models as innovation and renewal for incumbent firms (Johnson et al, 2008).

Overall, new business models can be seen as more customer centric (Mansfield & Fourie, 2004), taking on new forms of collaboration for value creation that necessitates a systems perspective (Seddon et al, 2004). It is also seen as a change in the unit of analysis from the firm to the value-creating system, which spans boundaries (Zott & Amit, 2010), and the need to focus on organisational activities that contribute to that system. This is the case with outcome-based contracts, which we shall consider next.

**Outcome-Based Contracts**
Traditional equipment-based service contracts consist of maintainence, repair or overhaul activities where the cost of replacement parts may or may not be included (Van Weele, 2002). Some are cost-plus contracts with detailed cost structures to ascertain reimbursement with a pre-determined profit percentage (Kim, Cohen & Netessine, 2007). More recently however, there have been an increasing number of contracts that centre on the outcomes of equipment instead of the resources required for its provision (Ng, Maull & Yip, 2009). For example, Rolls-Royce's 'power by the hour®' service to maintain engines is reimbursed based on how many hours the engine is in flight. Such outcome-based contracts aim to achieve necessary outcomes instead of a predetermined set of specifications or activities (Bramwell, 2003).

Outcome-based contracts (OBCs) theoretically change traditional business models in three ways. First, they ensure that both parties are aligned towards the incentives of the outcome. In traditional contracts, firms can be resistant to make voluntary and unilateral commitments outside of the contract, preferring expensive safeguards instead (Parkhe, 1993). OBCs create a mutual orientation structure capable of reducing opportunistic behaviour (Kale, Dyer & Singh, 2002), which indicates the ability to induce desired behaviours arising from the inducements within the contract, and therefore reducing the servicing cost for the customer over the longer term. Current literature indicates that with shared ownership of an outcome, both parties become 'mutual hostages' to the outcome, and so opportunism will likely decrease (Teece, Pisano & Shuen,1997).

Second, OBCs place the primary risk of outcome delivery on the firm, and secondarily on the customer. As the firm bears the greater proportion of the risk in achieving outcomes, it has the opportunity to integrate resources for value creation and value realisation by the customer (Madhok & Tallman, 1998), allowing the firm to earn better margins through more effective and efficient integration of the resources of both parties (Nooteboom, 1996; Dyer, 1997). Firms can therefore find in the longer term, that investing in the design of more reliable products and more efficient repair and logistics capabilities can increase profitability.

Third, achieving such a coordination role in OBCs enables the firm to fully master such a capability, which could allow it to increase its market share through further such contracts. The firm can be incentivised to make additional commitments outside of the contract terms, based upon the potential extraction of future revenues from such a capability. This would further increase the mutual orientation, and so results in OBCs being a self-enforcing agreement.

Some equipment-based service contracts are progressively becoming outcome-based, hoping to increase customer satisfaction, decrease costs, and reduce financial audits (Kim et al, 2007). This suggests that OBCs are a new business model that changes value drivers to partnered outcomes instead of billed activities (Demil & Lecocq, 2010). In doing so, it changes the focus from value capture to value co-creation (Hedman, 2003; Shafer et al, 2005), the dominant logic of 'selling to' to 'creating value with' the customer (Storbacka & Nenonen, 2009), and the unit of analysis from the organisation to the collaborative value-creating system (Zott & Amit, 2010).

Delivering on OBCs can be challenging (Ng & Nudurupati, 2010), requiring the firm to manage collaboration with customers. Business model literature suggests understanding changes in organisational activities (Zott & Amit, 2009). Furthermore, the fundamental theoretical issues supporting the dynamic firm-customer relationship in an OBC need to be considered. Literature in strategic alliance suggests the need to be able to cooperate and combine resources of both parties in the most effective and efficient manner (Gulati & Singh, 1998; Nickerson & Zenger, 2004). Conceptual and empirical studies in alliance literature have highlighted the challenges in achieving such coordination, including information sharing, cultural differences and conflict management (Das, 2000;

Dyer & Singh,1998: Reuer, Zollo & Singh, 2002).

We therefore propose that a successful change in business model to deliver on OBCs depends on developing the firm's capability to achieve cooperation with the customer as proposed by alliance literature. It also incorporates three key issues for the firm within the value-creating system: the **value** that is to be created, the **variety** that the system is subjected to and finally, how the firm could maintain **viability** from the new business model.

## Value

Scholars have described value as that which an individual derives from an offering due to the individual's ability to co-create that value with the offering to achieve his/her outcomes (Vargo & Lusch, 2004, 2008; Tuli, Kohli & Bharadwaj, 2007). Such value co-creation occurs through a process of an individual integrating his/her resources with the offering to achieve value. The co-creation of value is central to S-D Logic (Vargo & Lusch, 2004, 2008), which conceptualises service as the co-creation of value between the individual and the firm through an integration of resources accessible to both parties. It has therefore been proposed (Vargo & Lusch, 2004; 2008) that firms do not provide value, but value propositions, with value realised through co-creation interactions with the customer to achieve their goals. We argue that the value-in-use created through such interactions may not be all functional, but also emotional (Mattsson, 1992).

Understanding co-creation therefore requires the understanding of customer consumption processes (Ballantyne & Varey, 2006; Ng & Smith, 2012). Achieving value-in-use through co-creation has received considerable attention (Payne, Storbacka & Frow, 2008; Grönroos & Ravald, 2010; Sandström, Edvardsson, Kristensson & Magnusson, 2008; Heinonen & Strandvik, 2009), and most scholars have acknowledged that value-in-use is achieved in context. Since context is not completely certain, there is the potential for new experimental use to occur and the design of a product may not have accounted for different contexts of use, and so design for this beforehand can be challenging. This can be especially so when there are many contexts, i.e. contextual variety, which we shall discuss next.

## Variety

Given that value is created in a use situation, contextual conditions of that situation could affect its co-creation (for literature on situational and contextual value, see Beverland, Farrelly & Woodhatch, 2004; Flint, Woodruff & Gardial, 2002; Lemon, White & Winer, 2002; Lapierre, Tran-Khanh, & Skelling, 2008). Palmetier (2008) states that contextual variables may arise from changes in the physical environment, originating either from the provider and/or from the customer themselves. In any use of an offering, there could be a number of contextual factors affecting value creation, and such contextual factors will result in contextual variety in the way value is co-created, even by the same individual. This is particularly so for *continual* use of equipment over time. This is consistent with a systems perspective, where variety is the measure of the number of different states in a system. Consequently, variety is a measure of complexity as it counts the number of possible states of a system. *Contextual* variety as we describe here, is the number of different states in a system caused by different contexts of use.

It is when contexts begin to change more rapidly and not according to normal expected contexts of use that the degree of contextual variety increases. Thus, a high degree of contextual variety is an increase in the heterogeneity of the contexts that deviate from the most likely contexts of use for which the offering was originally designed. For example, research in manufacturing has shown that requirements gathering may not be able to understand, exhaustively, all the sets of possibilities surrounding customer requirements for the use of the asset (Potts & Hsi, 1997). Therefore, the implication is that every product is a manifestation of trade-offs between different sets of possibilities in contextual use, and the firm has to acknowledge that there will be some contextual variety that arises from the set of possibilities not taken into account, or not deemed to be feasible for the design and manufacture of the product. In this sense, therefore, service activities post-manufacturing can help manage unexpected contextual variety when it arises. However, the provision of service activities to enable value co-creation under high contextual variety can be costly to the firm, eventually threatening its viability. This suggests a need to design the asset for value co-creation under contextual variety in the first place, where possible, as not doing so may put the firm's viability at risk, an issue which will be further discussed next.

## Viability

Stafford Beer (1979, 1981, 1985) introduced the Viable Systems Model (VSM) to describe the necessary conditions for viability. Viability is defined as *the ability to maintain an independent existence within a specified environment*. In business, a viable firm is able to obtain funding or revenues for its offerings above the cost of delivering them. The management structure of the firm exists to support the process of profiting from its offering, without which it would become unviable.

The viable systems approach suggests that there are five systems necessary to ensure viability; this is illustrated in Table 1.

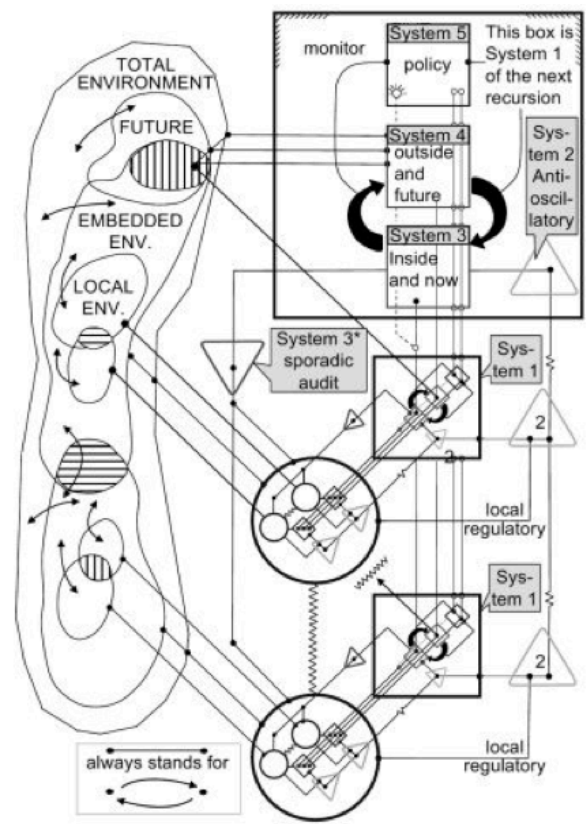

*Figure 1: A Viable System Model (source: Beer, 1984)*

System 1 (shown in Figure 1) is where the firm operates within an environment, depicted by a grey oval form. This system has to deliver despite changes in the environment, so it must have the capacity to adapt, cope and return the entity to stability. System 1, which is made up of the operations that justify the existence of the system (Beer, 1981), includes the management of these operations, but excludes senior management, which is considered as a set of services to System 1. Without System 1, there would be no reason for the firm to exist. A firm's environment consists of its customers, suppliers and regulators, which all could perturbate and disrupt the firm's core System 1 operations. Collectively, Systems 5/4/3 represent the meta system (future planning) and Systems 1/2/3 represent the current system (present planning), with System 3 as the key controlling bridge between the activities of Systems 1/2 and the management of Systems 4/5. To achieve *homeostasis*, i.e. self-regulation that maintains internal stability, the system requires *resources* and *management* (Golinelli, 2010). There are three main aggregate homeostats in the VSM (axioms of management):

- The homeostat in System 1 that stabilises the operations of the firm with its markets along the horizontal axis.

- The homeostat 3/4 maintaining System 3's coordination of the present with System 4's focus on the future.

- The homeostat that balances the horizontal variety between the System 1s and their environment and the vertical variety from Systems 1 to 5.

| SYSTEM | Description | Elaboration | Traditional company functions | Human body functions |
|---|---|---|---|---|
| 1 | **Key transformation** | This system has to deliver what it has been designed to do, despite changes in the environment, so it must have the capacity to adapt to be able to cope and return the entity to stability. A firm's environment consists of its customers, suppliers, regulators, all of which could experience perturbation which could disrupt the firm's core operations. | Operations Management – core value transformations. Recursions of viable systems | All the muscles and organs. The parts that actually DO something. The basic activities of the system. |
| 2 | **Conflict resolution, stability, coordination** | System 2 coordinates between the various recursions in System 1, so that common functions could be coordinated within the group efficiently. Note that System 2 is not autonomous, as none of the activities earn any revenues, although having an effective System 2 could save costs for the firm. | Account payable/ receivable IT support Health and Safety Travel Tax Compliance Administration | The sympathetic nervous system which monitors the muscles and organs and ensures that their interactions are kept stable. |
| 3 | **Internal regulation, optimisation, synergy.** | System 3 is the executive function of the group. The firm should be organised in such a way that the whole firm benefits, and even though some parts of the firm may not have the direct incentive to operate for the collective, System 3 ensures that they do, often leading to resource bargaining and lobbying. System 3 star is the part of System 3 that is required occasionally to enter System 1, often to cope with a crisis. System 3 star often includes internal audit, finance audit or compatibility audit where the purpose is not to micro-manage but to do a check to ensure System 1's effectiveness and agility. | Management accounting, production control. operations planning and control /audit – rules, resources, rights, responsibilities – interface between 4/5 and 1/2 | The Base Brain which oversees the entire complex of muscles and organs and optimises the internal environment. |
| | **Adaptation, dealing with a changing environment, forward planning.** | System 4's role is to scan the horizon, observe and forecast a future and plan for it. To do so, it must have a clear view of System 3 (current state) and where it needs to go to ensure survival. System 4 has ongoing conversations between its current state and its future state, setting up future resources and developing new offerings. Systems 3/4 homeostat is expected to maintain the tension between a future state and the current state. | Management, marketing, strategy, environment scanning (for adaptability) | The Mid Brain. The connection to the outside world through the senses. Future planning. Projections. Forecasting. |
| 5 | **Ultimate authority, policy, ground rules, identity.** | System 5's job is to maintain the System 3/4 homeostat, ensuring that the firm survives at present and remain viable for the future. System 5 also tackles the issue of the firm's identity and its mission. Much of business policy and strategic governance sits within System 5, which asks if the firm is doing the 'right' thing, rather than just doing it right. System 5 also manages the vertical variety of its own system from System 1 to 5, while balancing the horizontal variety between the systems and the environment. | Board of directors, business policy (decisions to maintain entity, balance demands from all parts, steer the organisation) | Higher brain functions. Formulation of Policy decisions. Identity. |

*Table 1: Beer's Viable Systems Model*

These three homeostats achieve stability in the firm to ensure its continued viability. It is important to note that the system in focus has to have a purpose. "Without a purpose, it is impossible to define a systems boundary. An essential basis for identifying and organising a system structure is to have a sharply and properly defined purpose" ([Forrester, 1968] as quoted in [Richardson, 1981]). The boundary of the system is an imaginary line separating what is inside from what is outside, for modelling purposes. This is important as the boundary specifies the scope of the system that achieves viability. Customer resources being 'outside' suggest that the firm has no systemic control over such resources, and information from the customer may be seen as 'perturbation' or 'disturbances' to the system. However, customer resources placed inside the system in focus suggest that the firm has some coordination or control capability.

The observation of system boundaries has many implications, including the potential for recursive behaviour within the levels (hierarchy) of systems. Recursion is essentially the process that an activity (procedure) goes through when one of the steps of the activity involves invoking the activity itself (often with a different set of parameters). This of course risks an endless loop, but recursion can be defined such that in certain cases (sets of parameters) the activity completes, no longer calling itself.

# VALUE, VARIETY AND VIABILITY: DESIGNING A VIABLE SYSTEM FOR VALUE CO-CREATION

The focus of this paper is to analyse a firm's System 1 operations as it moves from manufacturing to designing for value co-creation, where the value proposition changes from manufacturing an asset to the co-creation of outcomes in a combination of assets and human activities. Such a move transforms System 1's operational purpose from that of 'production' to 'achieving outcomes collaboratively'. The latter operations often result in the System 1 operation being a complex service system of people, processes, technologies and equipment. However, there is little understanding of what framework could inform the configuration of System 1 resources to achieve viability, whilst ensuring outcomes are achieved. Beer professes, "By finding invariances that underlie viability, is to make all of it susceptive to uniform description" (Beer, 1985). The notion of an *invariant,* i.e. a factor unaffected by the surrounding changes, is explored, and the purpose of this paper is to derive an invariant framework required for a manufacturer to achieve service transformation, ensuring viability to achieve outcomes with the customer.

As a firm moves from manufacturing an asset to offering outcomes, it immediately inherits the problem of contextual variety, as discussed earlier. Delivering an asset to customers for which they realise the value in their own time is quite different to promising them that their outcomes can be achieved collaboratively across the varied contexts. Achieving outcomes from variety of use is subject to the Law of Requisite Variety, which originates from the field of cybernetics, control and systems theory (Ashby, 1956); this essentially states that in active regulation only variety can destroy variety (Ashby, 1969). In other words, the more complex and variable a system becomes, the more flexibility and variety is required to manage those changes. This leads to the somewhat counter-intuitive observation that the regulator must have a sufficiently large variety of actions to ensure a sufficiently small variety of outcomes. Furthermore, it has important implications for practical situations; since the variety of perturbations a system can potentially be confronted with is unlimited, we should always try to maximise its internal variety to be optimally prepared for any foreseeable or unforeseeable contingency (Heylighen & Joslyn, 2001). Naturally, this has implications for systems of all types, including organisations, economies, families, interpersonal relationships and mental processes.

The Law of Requisite Variety was restated as only variety can absorb variety (Beer, 1979), because for a system to remain viable, variety must be managed. However, current literature does not provide any answers towards the resource configuration required within System 1 to manage that use variety and to successfully co-create value with the customer, where resources to co-create value are a combination of assets (equipment or goods) and human activities (people and processes). Indeed, most literature refer to the notion of 'servitization' as simply adding on service features (human activities) that relate to the core tangible asset to create additional exchange value, and consequently, boost revenues and the bottom line. There is very little literature that offers a framework to understand how value could be co-created to achieve outcomes when the value proposition is a combination of assets and people, within a system of processes and in combination with customer activities.

S-D Logic (Vargo & Lusch, 2004, 2008) proposes that "goods are a distribution mechanism for service provision" and that all offerings are services. While conceptually, it can be regarded that all offerings aim to achieve outcomes, it can be argued that the outcome achieved through an 'indirect service provision' (asset) requires more customer resource to realise than an outcome made possible through a firm's direct service activities, a point acknowledged by Vargo & Akaka (2009). In other words, assets are seen as enabling provisions while direct human activities are seen as relieving provisions (Normann, 2001). Furthermore, the capability to achieve the same outcomes whether through direct or indirect service provision requires a different set of capabilities from the firm. Neely (2008) provides empirical evidence that servitizing firms often generate lower profits as a percentage of revenues compared to pure manufacturing firms. Neely (2008) attributes this to the inevitable changes in value propositions that such a change to capability entails. This is echoed by many authors who continue to highlight the need to explore the transition from manufacturing to service (e.g. [Pawar, Beltagui & Riedel, 2009; Johnstone, Dainty & Wilkinson, 2009; MacDonald, Martinez & Wilson, 2009; Oliva & Kallenberg, 2003]). They recognise the need to explore the operational elements and to do so with a customer orientation (Johnstone et al, 2009), with many looking to S-D Logic as a lens through which this could be possible (Pawar et al, 2009; Macdonald et al, 2009). S-D Logic considers value co-creation as a process involving the integration of resources and recent research have empirically attempted to visualise how resources are integrated for value creation in OBC (see Ng, Parry, Smith, Maull & Briscoe, 2012). However, the resources for co-creation by the firm delivering an indirect provision, which in turn specify the capability of the firm, is clearly different from the resources for the same firm delivering service activities directly. From a viable systems perspective, if the resources to specify the core transformation of System 1 begin to change, creating instability, and if the management of System 1 fails to regulate to achieve homeostasis, the firm could quite quickly find itself becoming non-viable as evidenced by firms attempting to 'servitize'.

Consequently, in the new business model where a firm transitions from a manufacturer to an outcome-driven organiser of value creation, we are interested to discover the threats to viability and the drivers to direct or indirect service provision that ensure continued viability even while value, together with high contextual variety, is being co-created with the customer. This is the research question we ultimately seek to answer.

## METHODOLOGY

We consider three longitudinal case studies of three defence organisations who have contracted based upon outcomes. All three were awarded for the service of equipment (asset) they had originally sold to the customer. However, unlike conventional equipment-based service contracts where the firms are paid based on activities, repairs or spare parts used, these contracts were awarded on the basis of the *availability* of the equipment. The first organisation manufactures fastjets for the military, with the outcome being a 'bank of flying hours'. The second organisation manufactures missile systems with outcome being the availability of the system, while the third is an engine manufacturer providing the outcome of 'power by the hour'. The delivery of these contracts serves as an exemplar for complex service systems where both parties are focused on achieving outcomes; the firm's value proposition is co-produced with the customer (to achieve the outcomes); and the customer co-creates value with the firm through the *use* of the equipment. These service contracts were operating under complex relationships between clients and service providers and therefore relied heavily on both indirect service provision (e.g. tangible equipment) and direct service provision (e.g. knowledge and relationships through human resources) to deliver the outcome of the contract, through complex socio-technical systems management.

Case study research is useful when the aim of research is to answer "how" and "why" questions (Yin, 2003). Data for each case study was obtained through qualitative interviews, participant observations and company internal documents (Dooley, 2001). The logic behind using multiple methods is to secure an in-depth understanding of the case.

A total of 50 in-depth interviews were conducted with stakeholders from the firm and the customer over three years, to obtain a longitudinal understanding of the phenomenon. These interviews were audio recorded and subsequently transcribed, coded and categorised.

A qualitative approach was chosen, as a depth of understanding was required to analyse the way OBCs as a new business model were managed. Furthermore, the consideration that different parts of these large organisations may also potentially be at different stages of a transformation re-enforced the need to adopt a qualitative approach, to circumvent the risk that participants' particular social and institutional context may be lost where the collected data is quantified (Kaplan & Maxwell, 2005).

## FINDINGS

We found the **nature of emotional value to be co-created i.e. the customer experience**, the **degree of contextual variety** and **firm's 'legacy' viability** threatened the viability of the firm. To counter the viability threat, the firm uses **(a) Asset Provision for Scalability and Replicability of value proposition, (b) Service Provision for variety absorption and co-creating emotional value, and (c) Scalability and Absorptive Resources of the customer as an influential factor for its direct/indirect provisioning.**

## Threats to Viability

### Nature of Customer Experience to be Co-created

First, the nature of value to be co-created has an impact on the type of resources used in System 1. In all three cases, we found that the value consists of not only practical and logical dimensions (Mattson, 1992) (labelled jointly as *functional dimensions*) but also an emotional dimension in the form of the *experience*. In each of the cases, it was not only the functional dimension of value that was important to the customer, but the customer's perception had to be transformed into one that believes outcomes were achieved or achievable. In other words, System 1 not only had to transform materials and equipment to achieve the outcomes; the customer also had to be convinced that the process of doing so was culturally and adequately aligned with the needs of the customer organisation. This meant that previously, when the organisation had only to deliver an asset, System 1 was all about resources for transforming materials and equipment in a factory setting and handing it over to the customer, an indirect service provision. Yet, when the value to be co-created was outcome-based, customer perception of the experience became an important element of that value. The customer became concerned with both the process as well as the achievement of the outcomes, and the firm had to engage with the customer in a different manner and through different resources to ensure the perceptions/experiences were attained. This was often achieved through relationships:

> *"I don't think we put enough spending into how much relationship is worth as a business. We tend to focus heavily on the things that you can touch and feel like erm somebody can write you a process or a procedure but it's the softer issues that make these things work the softer skills, the you know the way in which people interact, the way in which we operate with our customer once we are on his [site]. You know they are the things that really grease the wheels….that's the glue that makes all this work."*

This leads to our first proposition.

***Proposition 1: In co-creating value for customer experience, System 1 for the firm has to include the transformation of the customer to ensure viability***

### Degree of Contextual Variety
Second, the degree of contextual variety also had an impact on what resources were used in System 1. We found that contextual variety arises not merely from the context of usage, but in the moral hazard from equipment use when there is no sense of ownership. As one respondent puts it:

> " … it's like a car isn't it, you-know? I drive my car and abuse my car, whereas my partner looks after her car, so that gives different demands on the garage. …..If they don't do that in a logical way, following the process that's outlined in the manual – the data that we get back that we need to analyse to try and reduce [problems] on the [asset] and reduce the number of faults on the [asset] is flawed."

The variety of use became a serious issue as contracts required constant amendment to accommodate increasing sets of possibilities:

> "….The other thing of course is the contract doesn't stay the same, its constantly being changed and then the [outcomes] have changed they are going to want to give you extra work or extra scope so more and more things are coming into the contract and we go oh this is an amendment is that a purely fixed amendment is it variable is a mixture is it, so the baseline changes constantly as we move forward"

Our study found that contextual variety threatens viability in two ways. The first threat is from the firm being unable to absorb variety. This means that System 1 has not got the requisite variety to absorb contextual variety from use, and implies that the customer may be unhappy due to the firm's inability to accommodate certain contexts of use. This inflexibility threatens the long-term viability of the firm as it struggles to meet customer expectations in a timely manner, and it may find itself losing the customer as a result of that failure. The second threat is from absorbing too much variety, which disrupts the firm's internal system, challenging homeostatis. We found that when the contextual variety of use is high, the firm amplifies its variety through greater responses, and System 1 suffers the strain as inadequate resources are provided to stabilise the system.

***Proposition 2: In co-creating value for outcomes, the firm has to balance the attenuation and amplification of internal responses to match contextual variety to ensure viability***

**'Legacy' Viability of the Firm**

Our study found that when System 1 was operating purely as a manufacturer, it did not have to manage much variety. The firm's established viability was based on a transfer of asset ownership and when called upon, undertake maintenance and service activities, relegating the variety issue to a scheduling problem. However, when the firm is tasked to co-create for outcomes, it has to take responsibility for the outcomes within the customers' use situations, which results in the firm having to take proactive initiatives that are uncertain and where the absorption of variety may require different resources. It also meant that the transfer of responsibility requires the firm to be involved in customer contexts and use situations so as to obtain the benefit of reduced costs and reduced variety. Yet the following quote shows how this threatens the established system and challenges the mindset:

> "when I report back into mothership they would say, 'why are you worried about …the user? That's not the contract – you've just got to deliver the [outcome]'. And I'm saying, 'well hang on a minute…….why wouldn't you get closer to them? Because, in most cases, it creates a win-win situation where you're involved in terms of what the customer finally gets and, in financial terms, we gain anyway……but I'm struggling to get the back-end of the company to get that?"

Since the asset is now the responsibility of the manufacturer to achieve outcomes,

the co-creation activity no longer interacts in the same way as when the asset was the responsibility of the customer. Yet, System 3 could be controlling Systems 1/2 in a 'legacy' manner, while Systems 1/2 are struggling to cope with a different kind of variety entering the system. This leads to an imbalance:

> "I've got somebody sat in the back office at ….. who's just got it in his tray, having a cup of tea and thinking in weeks, months and years, when I'm trying to think in seconds, minutes and hours ….So that means back office needs to change the way they're organised and the way they work and what they're roles and responsibilities are and, in some cases, their capability as well."

**Proposition 3: In ensuring viability, the firm has to ensure that resources allocated to Systems 1/2 are in line with Systems 1/2 key operational elements and not legacy operational elements**

Our findings suggest that that the choice between indirect (modifying the asset) and direct provision (human interventions) interacted severely, and there is tension between resources for scalability and replicability (assets) and resources for variety absorption (autonomy, empowerment and human skills) to achieve outcomes. They also show that the choices of direct and indirect provisions improved the viability of the firm in different ways.

## Ensuring Viability in the New Business Model

### Indirect Service Provision for Scalability and Replicability of The Value Proposition

Our findings suggest that when firms were manufacturers, their viability came from production and transfer of ownership, which could be scaled in line with demand. In co-creating outcomes however, firms became increasingly challenged in scaling or replicating for growth due to embedded human capability.

> "…and service thing is not easy with this new model…we could get a different person and it won't turn out the same……and then there so many changes that you can't really design anything …the customer wants different things, solve different problems … there's a fire fighting mentality…"

Our findings show that high indirect service provision within a firm's outcome-based value proposition delivered low margins on a contract for two reasons. First, it makes *the system* less replicable because embedded human capability, particularly when skills and knowledge form a valuable resource, is not as easily transferable to other employees as assets are. This results in slower growth since systemic capability to achieve outcomes takes longer to acquire. Second, the human resource component makes the system less scalable. Whilst an asset could be scaled by increasing production lines and/or improving manufacturing capacity, complex service systems of direct and indirect provision are less easily scaled, resulting in investment or costs for a small project similar to that of a large project. Economies of scale are therefore harder to achieve in outcome-based environments.

To counter the above challenge, our study found that the firms became willing to change indirect service provision to achieve outcomes that could be more scalable and replicable, modifying the asset through redesign or incorporating technology insertions:

*"I think we're achieving better outcomes with the current equipment because we're starting to collect more [electronic health monitoring] data about what's happening; we're starting to have different discussions with the customer about what's happening so we can actually get a better understanding of what's happening and look for failures, or signs of failures happening before they actually fail."*

***Proposition 4: Redesigning and modifying indirect service provision (asset modification) ensures viability through scalability and replicability***

**Direct Service Provision for Variety Absorption and Co-creating Emotional Value and Experience**

Conversely, our study found that the use of direct service provision was essential to absorb contextual variety.

*"You then see that he can then use those relationships to either just sort of oil the wheels altogether speed things up or he could have a conversation say with the [customer employee] ……. he would talk to [person] and [person] would go and do it and at the end of the day the [customer employee] work for him so there is all that sort of complexity of relationship building and then you just know you are going to get benefit from that but things happen, things are much easier, things get smoothed through that could otherwise could become an huge issue."*

Human resources were used to absorb the impact of variety into the firm. First, in direct engagement with the customer, the firm would try to ensure low contextual variety by monitoring and engaging the customer on use behaviour:

*"So what it's driven us to do is start to focus more on managing [problems] and to do that we need to get closer to the user…. What are you doing with it? How are you [using] it … Erm, how are you looking after it? How are you doing your diagnostics? Are you in a maintenance policy with the level of maintenance that you're doing. Erm, start to look at the [user] and navigate his report in more detail. So we're gathering more and more data and starting to analyse that data and then coming up with solutions on how we might reduce the [faults]…. And then you get a win-win obviously, because that saves us money and it gives more [asset availability] to the end user. So that, predominantly, is what we aim to do – that support for [users] more than probably the contract would have wanted us to."*

Second, where the customer could do no more, human activities within the firm bridge the gap, albeit with some difficulties:

*"Now you can either spend two years having the fight and whinging or if you have got the relationships you can just, it will get sorted out so….. it just makes everybody's life a lot easier and things just get done."*

Thus, human resources through direct service provision amplify the variety being managed through responses to the customer, absorbing variety, i.e. human resources create responses that exhibit requisite variety.

Our study also found that human activities were instrumental in co-creating the experience. The firms had to design, within the service system, methods of how individuals' perceptions within the customer organisation were also

'transformed' as part of System 1 operations, i.e. management of the customer experience. The method varied across organisations. One used technological resources to allow the customer to 'view' the way they worked to create transparency and closeness, while the two other firms provided regular updates, even when not contractually required. All three organisations used relationships so that the customer 'perceived' the contract was in good hands and outcomes were on track.

> *"we're starting to have visual and verbal contact with the people that need to be helping us sort it – so they're starting to become part of it – they're starting to feel it.... it's about us understanding what we're actually delivering and changing our culture, environment, abilities and roles and responsibilities are aligned to it [the customer]"*

> *"I think they trust us; trust us to deliver excellence actually isn't a bad logo for somebody. I think they do trust us; they do know we know what we're talking about. We're excellent at fire-fighting – we're well known for that .... If there's a problem we are the world's best at solving them because that's interesting to us because that's our culture, you-know, we will throw people at issues… And to be quite honest we reward it as well; we reward people for sorting problems out for us."*

***Proposition 5a: Direct service provision ensures viability through absorption of contextual variety and co-creating emotional value and experiences***

Our study also found that contextual variety was a manifestation of latent demand, and that the variety of use belies the need for additional provisions from which the firm, if it provided them, could derive greater revenues:

> *"we get into an argument with the [user] that, …. they say 'the outcome isn't what we expected'. Now actually the outcome is what is expected but it's not what they now want because they want more......then what the user wants in terms of [outcome] is more than we've agreed...but it looks like it's going to improve [the] order book position"*

***Proposition 5b: Contextual variety provided an opportunity for firms to innovate and derive new revenues to satisfy customer latent need.***

**Interaction of Direct and Indirect Service Provision**

Our study found that the firm has to rethink its resources and how System 1 is configured for achieving outcomes, which could be different from how it was originally set up to manufacture and transfer the ownership of assets.

With the change of System 1 transformation activities from manufacturing to achieving outcomes comes a change in resources required to achieve that co-creation; this in turn comes with the challenge of whether the asset was designed correctly to support such activities. Our study found that an asset designed and engineered for a transfer of ownership to the customer so that the customer achieves the outcomes on their own, may not be the most optimal asset for delivering outcomes together with the customer, where such outcomes could be a responsibility of the firm.

> *"A classic example for me with the [asset], it was designed to be stripped and rebuilt in [our factory]. If we'd done that [at client location] it would have been designed*

*differently because we would have taken it apart differently, because [in the factory], we don't have to worry about [shelters to protect the assets] and all those sorts of things ……So there are parameters placed on you which the customer has to deal with in a [use] environment…and you need to now deal with that (when you are delivering outcomes)."*

Our study found that achieving outcomes began with the firm 'wrapping' human activities around an asset, without any serious thought about (a) the outcomes the system aims to achieve; (b) the resource combination of direct and indirect service provision to achieve the same outcomes; and (c) the business model that renders the system viable. Over time, the firms came to the realisation that the asset was not a "sacred cow" and the better it could absorb contextual variety of use, the less its dependency on human activities to absorb the variety and the better it could scale and replicate the system across contracts. Concurrently, the firms also became aware that understanding where contextual variety is highest and deploying human activities to absorb variety (either by attenuating or amplifying it) resulted in better engagement, higher satisfaction, and the co-creation of emotional and perceptual value in the customer experience. This is evidenced by the following quote from one of the employees of the firm when discussing their customer:

*"If there's a problem we are the world's best at solving them because that's interesting to us because that's our culture, you-know, we will throw people at issues… I think they do trust us; they do know we know what we're talking about. We're excellent at fire-fighting – we're well known for that ……"*

With the absorption of variety, co-creating customer experience through human resources and achieving scalability/replicability through assets, the firms started putting in place processes where contextual variety became a conduit for feedback on the degree of substitutability for indirect and direct provision for co-created outcomes, and also to drive both direct and indirect service innovation:

*"As we're starting to collect more data about how the customer uses them, either electronically – so does he know we're getting them? He knows we're getting it but he's happy for us to get that – or via interviews with [users] and those things – it's helping us understand better to look for trends; to look for potential failings of those mechanisms so that we can then, a) stop it happening but also look at that particular area and say, 'well, would we do that differently?"*

***Proposition 6: Scalability and Replicability of Direct Service Provision (people and processes) are dependent on the design of the indirect service provision (asset) for variety absorption***

**Scalability and Absorptive Resources of the Customer for Value Co-creation**
Our study also found that the degree of skills and knowledge for the customer to realise and co-create value interacted directly with both direct and indirect service provisions. Assets which are better platforms for co-creation, better able to absorb greater variety, either through modularity or clever design, required lower skills and knowledge from customer employees, and less of such resources. This implies that the scalability and replicability of the provider service provisioning may not merely lie with the firm's direct and indirect service provisions, but with the resources required on the customer side to realise the provisions for outcomes. Conversely, complex assets that had greater technological capabilities required more complex sets of resources to use and operate them. This in turn had an influence on the firm's choice of direct or indirect service provision.

> "if you look at a lot of the land equipment ... So to take the average lorry that was used by the Army, it was used ... you needed to know how to take engines apart and you'd have to change wheels, you now need almost a degree in Electronics because the whole thing is now computerised so, in a sense, they've actually created a problem there, where at one time running a tank or a lorry was quite cheap, you actually now have to change the type of person who now actually manages that because the average sort-of mechanical person can pick out and can do that – it doesn't get fixed any more......in the past where their Army recruits came in at basic mechanic, 'can you undo that bolt?' they're actually having to come in at graduate level to actually be able to manage and understand the complexity of the equipment they're now getting. "

Customer resources for co-creation therefore had four types of impact on the firm's service provision. First, the more complex indirect service provision would require more complex customer resources to co-create value. Second, the customer activities to realise and co-create value with the indirect service provision could be more replicable and scalable if the asset was easy to use, providing efficiency gains to the customer. This also meant that the firm's direct service provision became less complex, because the customer required less support. Third, if the asset could absorb greater contextual variety, the customer would know what to do in different use situations and so less use variety permeates into the firm's system, requiring less direct service provision to absorb the variety. Fourth, customer resources themselves could absorb contextual variety by deploying their own internal resources so that the environment is less disruptive on the provider's system.

***Proposition 7: Customer resource requirement to co-create value in contextual variety changes the nature of direct and indirect service provision by the firm and vice versa***

## DISCUSSION

### Value, Variety and Viability - Extending Service Dominant Logic for the new business model of OBC

To achieve co-created value-in-use that could be for both functional outcomes and customer experience in OBC, our study found that direct and indirect service provision interacted with customer activities to realise the offerings. Also, the configuration depended on the value to be co-created, contextual variety that needed to be absorbed, as well as the need for viability of the provider.

Our findings suggest that four interactions exist in the co-creation system, as summarised in Figure 2.

Interaction 1: Increasing Scalability and Replicability means redeploying resources to indirect service provisioning

Interaction 2: Increasing Variety absorption and co-creating customer experience means deploying resources to direct service provisioning

Interaction 3: Customer activities that co-create value under contextual variety changes the nature of direct and indirect

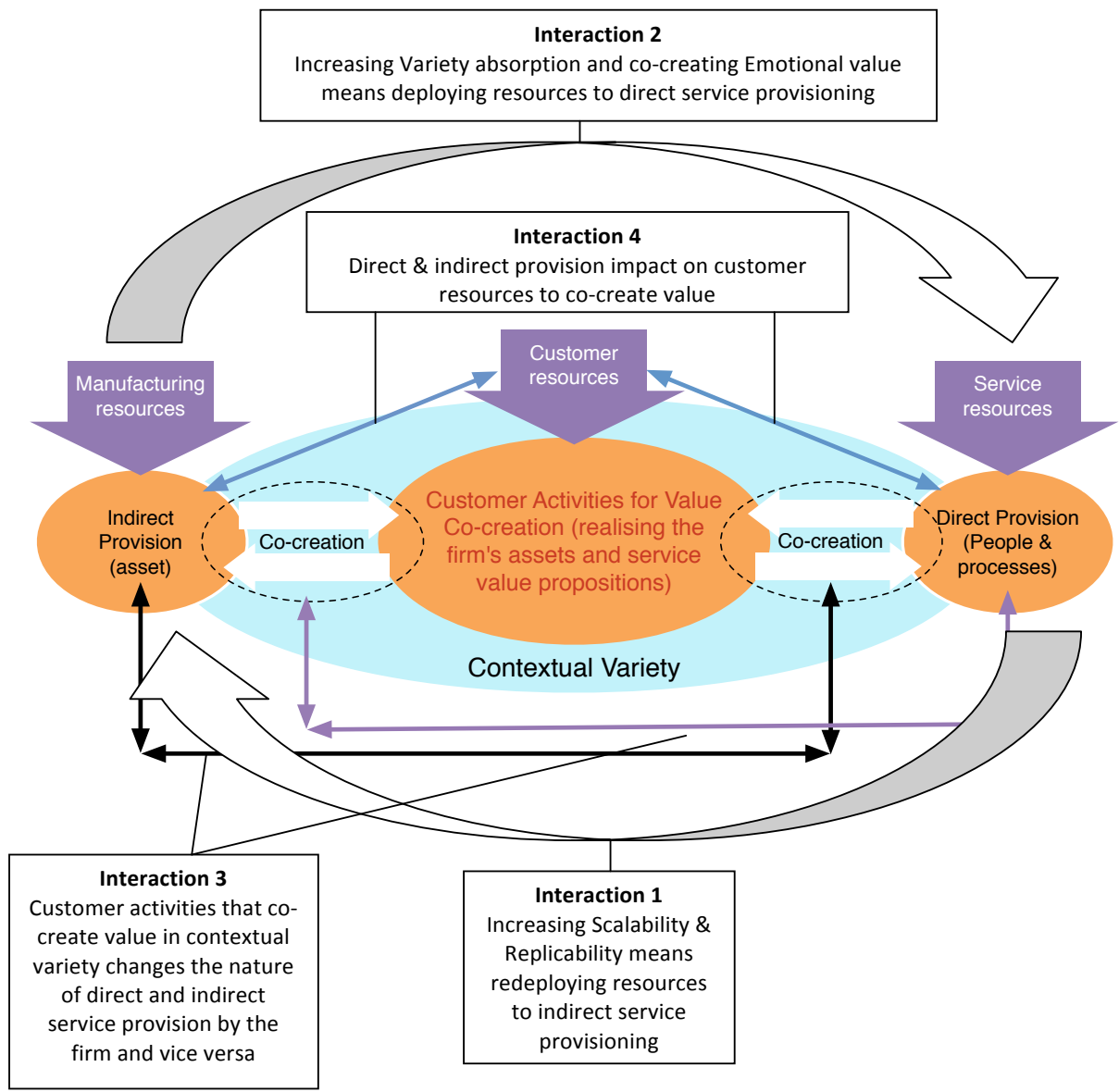

*Figure 2: Interactions Between Customer Resources and Activities and the Firm's Direct and Indirect Service Provision in a System of Value Co-creation*

service provision by the firm and vice versa

Interaction 4: Direct and indirect provision impacts on customer resources to co-create value.

Our study showed that the difficulty in the change of business model may lie not merely in the activities of service personnel, or in processes that surround the asset, but in the design and engineering of the asset itself to support activities of service personnel in combination with customer resources. Consequently, if the asset was originally designed towards a different set of boundaries i.e. the firm is only responsible until the ownership was transferred, it may need to be redesigned with this new set of boundaries where both are now responsible for co-created outcomes.

The firm's value proposition for co-created outcomes consists of both direct (human activities) and indirect (asset) service provision, and the tension between them that threatens viability lies in the degree of replicability and scalability. Our study found that direct service provision challenges the viability of the firm through its inability to scale for growth and replicate across other contracts. The findings indicate that customer-facing teams held the knowledge of the customer, their contexts and their demands within human capability and skills, to the extent that although service to the customer was excellent, every contract became a new design, a new team and a new set of relationships. To reduce the risk to viability, firms have looked into the redesign of the asset. Yet, we found that direct service provision absorbed contextual variety and co-created customer experiences, leading to better customer engagement and experience. In addition, contextual variety was a manifestation of latent demand and new markets, and innovation could arise when variety of use is closely monitored.

Our findings suggest a paradox in that as indirect service provision (assets) become more technologically capable and complex, which could increase its exchange value to the firm, both the direct service provision (human activities) and the customer resources (resources to co-create value) become less scalable and replicable (and in many cases, more expensive). This in turn could result in an inability in the overall co-creating system to achieve outcomes in a scalable and replicable manner, which may threaten the viability of the firm in the long term. From a business model perspective, the risk of higher co-creating resources by the customer may compel more contracts based on outcomes, which could reduce customer co-creating resources, but may result in the firm re-engineering the asset to enable better use capabilities for contextual variety.

## A Proposed Viable System of Indirect and Direct Service Provision With Customer Activities for the New Business Model of OBC

Our study suggests that the new business model of co-creating functional outcomes and customer experience consists of three main System 1 operational elements that interact: That of transforming indirect service provision (materials and equipment), transforming direct service provision (people, information and processes) and transforming the customer employees, as shown in Figure 3. The connections between these System 1 entities are closely coupled, resulting in emergent effects. Serving the three entities are resources accessible by System 2, which consists of a regulatory centre for each element of System 1, and an overseeing regulation at the senior management level. System 2 plays a crucial role in achieving outcomes as it serves not only to regulate the interactions between elements of System 1, but also functions as the most stable and efficient configuration of direct and indirect provision to achieve customer transformation and co-creation within some level of contextual variety. System 2 is therefore tasked with balancing scalability and replicability with variety amplification and attenuation within System 1. To co-create value with customers, System 2 also achieves an important regulatory function; where the firm is unable to amplify variety to match customer's contextual variety, System 2 has to be able to harness *customer resources* to reduce variety in the system, through changes of customer use behaviours achieved through social resources such as relationships and culture. Beer (1984) considers this role as the 'damping of oscillations'.

The viability of a firm transforming from a manufacturing concern into a service organisation co-creating valued outcomes therefore concretely implies the following changes to the business model of the firm:

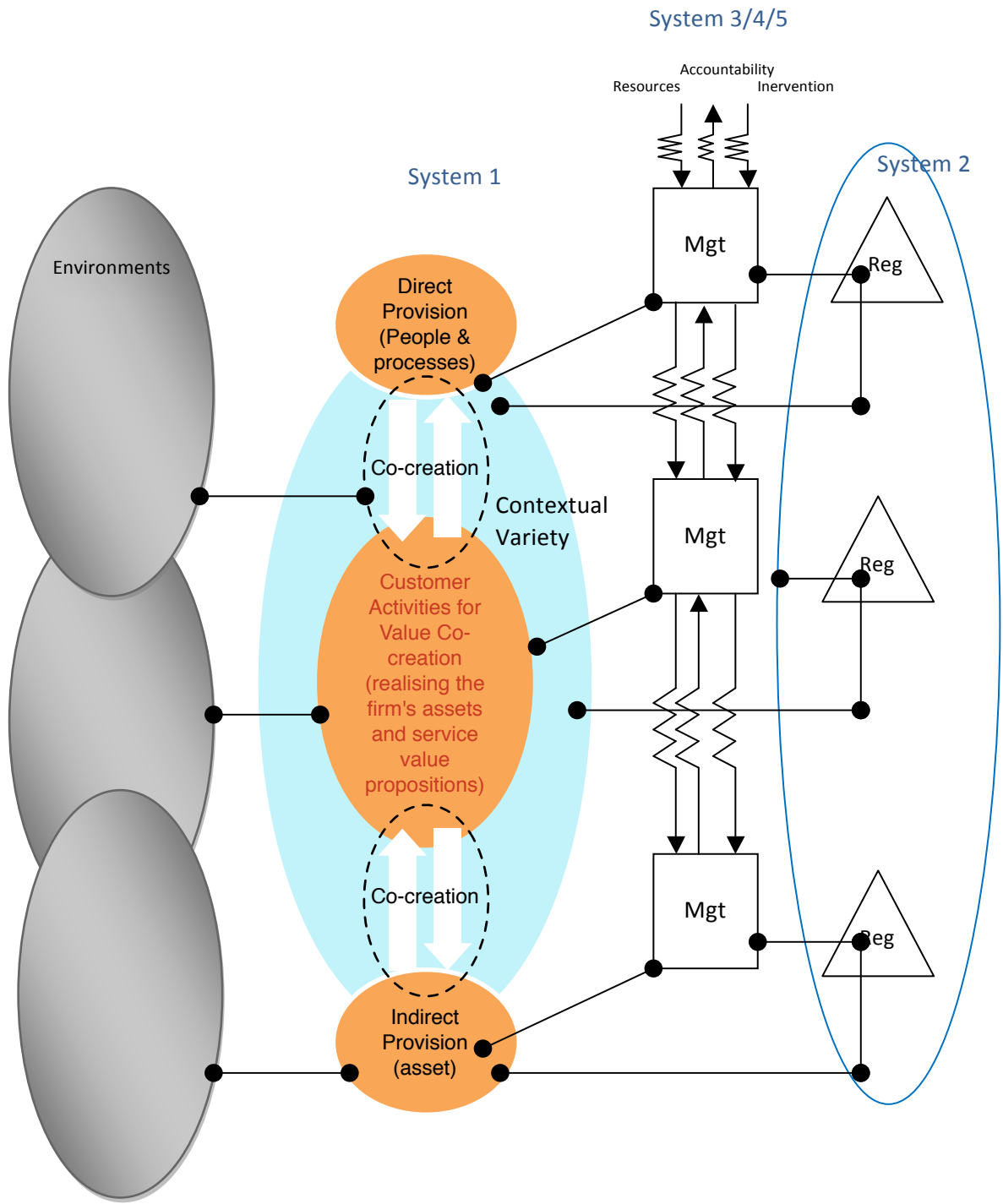

*Figure 3: A Viable System for an Organisation Co-Creating Outcome-Based Value In Use*

1. The redrawing of system boundaries to include the customer within its boundaries, but which *must also include* Systems 3 and 2's capability to harness customer resources to amplify or attenuate variety in the system caused by uncertain environmental factors;

2. The additional System 1 element that *transforms customer employees for a positive customer experience* in addition to transforming indirect service provision (design and manufacturing of asset) and direct service provision (design and implementation of people and processes);

3. The customer transformation operational element could be interventionistic on the customer's co-creating activities at higher level of recursion, over which the firm may not have control ;

4. A more tightly coupled System 1 operational entities where transforming indirect service provision (design and manufacturing of asset) for value co-creation with the customer interacts with transforming direct service provision (design and implementation of people and processes) as well as with customer co-creation activities. A tightly coupled System 1 creates emergent effects embedded within the customer experience;

5. System 2's ability to coordinate between the three operational entities through allocation of different resources required for scalability/replicability and variety amplification/attenuation through redesign of direct or indirect service provision over time; and

6. The support from Systems 3, 4 and 5 that could also be collaborative in nature with the customer to allocate resources and control the overall system.

## CONCLUSION

Beer's (1979) first axiom of management suggest that the sum of horizontal variety disposed by all the operational elements must be equal to the sum of vertical variety disposed by the six vertical components of corporate cohesion.

Our study suggests that organisations structured around manufacturing require a re-evaluation of operational elements and viability within the system when they adopt OBCs, transforming towards a new business model of value co-creation under contextual variety. Homeostasis could be seriously disrupted by high contextual variety if they are not able to do so, and the viability of the system would be threatened. We propose that understanding value-in-use, contextual variety, and a system's perspective of viability are the **three core principles** for the new business model in OBC that is able to co-create value with customers through both direct and indirect service provision.

The benefits of our approach include extending the work of S-D Logic. Specifically, operand and operant resources, in the context of value co-creation, is formed from direct and indirect service provision of the firm together with customer activities to realise the offerings in context. Therefore, our efforts provide greater understanding of value co-creation in complex equipment-based systems, including a discussion on the firm's viability as it invests in such capabilities. The limitations of our approach centre around the need for a larger study to confirm wider applicability of the understanding we have gained, so that wider conclusions can be drawn with regards to the emerging design of the business model observed.

Goods are often designed purely within the domain of engineering and product design, often placing human activity in service as a supporting role to the equipment. Our study considers the design of both equipment and human activities, without privileging either entity, for the purpose of co-creation with the customer in a complex service system. Our work contributes to the understanding of the interface between equipment (assets) and human activity, as direct and indirect service provision for new business models of OBC aimed at co-creating value with customers.